\documentclass[twocolumn]{autart}    % Enable this line and disable the 
\usepackage{graphicx}          % Include this line if your 
\usepackage{epsfig}
\usepackage{subfigure}
\usepackage{amsmath}
\usepackage{amsfonts}
\usepackage{amssymb} 
\usepackage{epstopdf}
\usepackage{subfigure}
\usepackage{xcolor}

\begin{document}

\begin{frontmatter}

\title{A uniform reaching phase strategy in adaptive sliding mode control \thanksref{footnoteinfo}} % Title, preferably not more 

\thanks[footnoteinfo]{The material in this paper was not presented at any conference. }

\author[a]{Christopher~D.~Cruz-Ancona}\ead{cdiegoca89@gmail.com},    % Add the           % e-mail address 
\author[a]{Leonid~Fridman\thanksref{coraut}}\ead{lfridman@unam.mx},  % (ead) as shown
\author[b]{Hussein~Obeid}\ead{hussein.obeid@univ-fcomte.fr},
\author[b]{Salah~Laghrouche}\ead{salah.laghrouche@utbm.fr}
%\author[c]{Yacine Chitour}\ead{yacine.chitour@l2s.centralesupelec.fr}

\address[a]{Facultad de Ingenier\'ia, Universidad Nacional Aut\'onoma de M\'exico (UNAM), Ciudad de M\'exico 04510, M\'exico}  % Please supply                                              
\address[b]{Femto-ST UMR CNRS. Univ. Bourgogne Franche-Comt\'e/UTBM, 90010, Belfort, France}             % full addresses here.
%\address[c]{Laboratoire des Signaux et Syst\`emes, Universit\'e Paris-Sud, Centrale Supelec, Gif-sur-Yvette, Paris, France}
\thanks[coraut]{Corresponding author}

\begin{keyword}                           % Five to ten keywords,  
 Reaching phase; sliding mode control; adaptive control; barrier functions.         % chosen from the IFAC 
\end{keyword}                             % keyword list or with the 
                                          % help of the Automatica 
                                          % keyword wizard

\begin{abstract}                          % Abstract of not more than 200 words.
In adaptive sliding mode control methods, an updating gain strategy associated with finite-time convergence to the sliding set is essential to deal with matched bounded perturbations with unknown upper-bound. However, the estimation of the finite time of any adaptive design is a complicated task since it depends not only on the upper-bound of unknown perturbation but also on the size of initial conditions. This brief proposes a uniform adaptive reaching phase strategy (ARPS) within a predefined reaching-time.  Moreover, as a case of study, the barrier function approach is extended for perturbed MIMO systems with uncertain control matrix. The usage of proposed ARPS in the MIMO case solves simultaneously two issues: giving a uniform reaching phase with a predefined reaching-time and adapting to the perturbation norm while in a predefined vicinity of the sliding manifold.

\end{abstract}

\end{frontmatter}

\section{Introduction}
Adaptive sliding mode control (ASMC) is an efficient technique for compensating matched perturbations: uncertainties and disturbances without knowing their upper-bound \cite{bartolini2013,hsu2019,incremona2016,negrete2016,obeid18,oliveira2017,plestan2010,xiong2021adaptive}. ASMC should \textit{simultaneously} solve two issues: 
\begin{itemize}
\item [(i)]Reaching phase (RP). The controller's gain increases to a value confining the system's trajectories inside some neighborhood of a sliding set (NSS) in a finite reaching-time (RT).
\item [(ii)] Adaptive phase (ASP). Once in the NSS, the controller's gain is updated at the RT moment to maintain the system's trajectories following sliding dynamics.
\end{itemize}\vspace{-2mm} Whether all approaches above accomplish (ii) by keeping some NSS while fixing \cite{negrete2016,incremona2016} or reducing \cite{shtessel2012,plestan2010} the gains; or by ensuring a predefined NSS that will never be exceeded via barrier \cite{obeid18,obeid20,laghrouche21}  and monitoring \cite{hsu2019,oliveira2017} functions based gains. Their common characteristic relies on (monotonically) increasing the controller's gain to solve (i). However, even when RP theoretically occurs in a finite-time,  a dichotomy exists between the estimated RT and the size of initial conditions with the unknown upper bound of perturbations.  Therefore RT is unpredictable since the perturbation might not attain its upper-bound at the end of the RP as initial conditions could take any value.  The next example illustrates the compromise between the initial conditions and the perturbations' upper-bound with the  RT. \vspace{-4pt}
\subsection{Motivating example}\label{subsec:motexmp}
\begin{figure}[h!]
		\centering
		\includegraphics[scale=0.12]{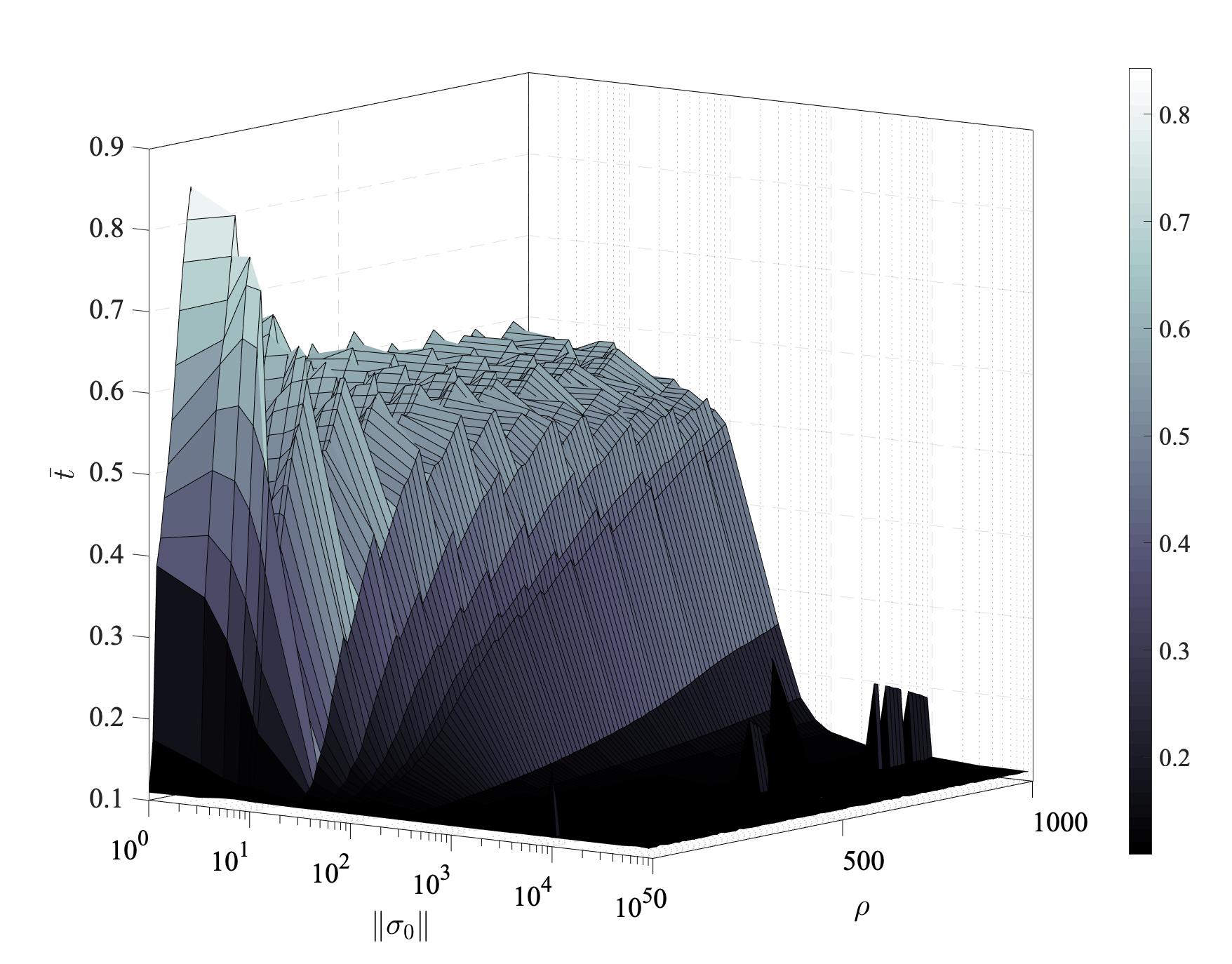}%{figs/BFSMC_rho10_eps05_plusOBEID.pdf}
		\caption{Non-uniform RT under increasing norm of initial conditions and size of perturbations}% with $\min \bar{t}=0.1099$}
		\label{fig:obeid_BFBSMC}
	\end{figure}

\vspace{-5pt}
Consider a system of the form $\dot{\sigma}(t)=H(t,\sigma(t))\nu(t) + f(t,\sigma(t),\rho), \:
\sigma(0)=\tfrac{b}{\sqrt{2}}(
10^{n} \: -10^{n})^T,$ with \vspace{-5pt}
\begin{equation*}
\begin{aligned}
 f(t,\sigma,\rho)&\!\!=\!\!\rho\begin{pmatrix}
a_1 + 0.4  \sin (\omega_1 t)+ 0.01  \cos(20t+\sigma_2)\\
b_1 + 0.2  \sin (\omega_2 t)+ 0.02  \cos(15t+\sigma_2)\\
\end{pmatrix},
\end{aligned}
\end{equation*}
\begin{equation*}
\begin{aligned}
H(t,\sigma)&\!\!=\!\!\begin{pmatrix}
1+\tfrac{1}{2}\cos(\sigma_1) & \tfrac{13}{30}\cos(\sigma_1)-\tfrac{1}{30}\sin(5t+\sigma_2)\\
0 &\!\!\!\!\!\!\! 1+\tfrac{1}{5}\cos(\sigma_1)+\tfrac{1}{10}\sin(5t+\sigma_2)
\end{pmatrix},
\end{aligned}
\end{equation*} where $H$ and $f$ denote the input matrix and a matched disturbance, respectively. Let $\rho, \:b,\:a_1,\:b_1,\: \omega_1,\: \omega_2\in \mathbb{R}_+$ and $n\in \mathbb{N}$ denote constant values parametrizing the perturbation and initial conditions of the system.  Consider the control input $\nu(t)=\hat{k}(t,\sigma)\sigma/\Vert \sigma \Vert$ during RP. Adopting an ASMC strategy (\textit{cf. } with \cite{obeid18,plestan2010}) of the form $\dot{\hat{k}}=\bar{K}\Vert \sigma \Vert,\: \hat{k}(0)=\hat{k}_{0}$, it can be ensured that for any $\varepsilon>0$ and $\bar{K}\in \mathbb{R}_+$ there exist $\bar{t}(\sigma_0,\rho)$ such that $\Vert \sigma(t)\Vert = \frac{\varepsilon}{2}$ at $t=\bar{t}$.  For simulation purposes, set $a_1=1$, $b_1=1.2$, $\omega_1=3$, $\omega_2=2$, $\bar{K}=100$, $\varepsilon=0.05$, $\hat{k}_{0}=0$ and take fixed parameters $\rho \in [0,1000]$, $n\in [1,4]$, $b\in [1,9]$ increasing the  upper-bound of the matched disturbance $\Vert f(t,\rho) \Vert_2 \leq 2.63 \rho$ and the norm of initial condition $\Vert\sigma_0\Vert=b \times 10^n$ in different simulations scenarios.  Fig. \ref{fig:obeid_BFBSMC} illustrates that even though RT $\bar{t}(\sigma_0,\:\rho)$ to a set $\left\lbrace \Vert \sigma \Vert =\varepsilon/2 \right\rbrace$ is finite, it is not uniform with respect to the initial conditions and the size of perturbation.% as $n,b,\rho$ take different values.  
\vspace{-10pt}
	
Moreover, if an expression for $\bar{t}(\sigma_0,\:\rho)$ is available, it is impossible to estimate it since $\bar{t}$ depends not only on the initial conditions but on  \textit{a priori }unknown upper-bound of the perturbation. An adaptive controller design based on the latter methods is unreliable since RT $\bar{t}$ to some NSS cannot be \textit{a priori} known nor to be estimated. 
\subsection{Contribution of the paper}
The paper presents a uniform adaptive reaching phase strategy (ARPS) ensuring a predefined convergence time to the predefined NSS, which can be useful for different ASMC algorithms \cite{negrete2016,obeid18,plestan2010} dealing with bounded perturbations with unknown upper-bound.  Similarly with \cite{chitour2020,ferrara2020,gomez2020,holloway2019,jimenez2020,song2017}, the gain of proposed controller is growing when the trajectories are tending to the sliding set.  But in contrast with a method \cite{chitour2020,ferrara2020,holloway2019,jimenez2020,song2017} the proposed controllers' gain is updating its value to the size of perturbation and kept bounded because the proposed method just requires the system's trajectories to reach first time NSS but not the sliding set. Therefore, the RT for the system's trajectories, starting from any initial conditions to reach NSS, is uniformly bounded by a predefined time constant despite of the presence of perturbations with unknown upper-bound.  At the first time moment when the trajectories are reaching the subset $\left\lbrace ||\sigma||=\varepsilon/2 \right\rbrace$, one of the ASMC from \cite{negrete2016,obeid18,plestan2010} should be switched on.

To show the efficiency of proposed APRS the barrier function (BF) approach is generalized covering two important classes of systems: MIMO systems and systems with uncertain control matrix with unknown upper-bound. Then it is shown that a combination of proposed ARPS with BF adaptation ensure that: convergence to NSS is given in a predefined-time; the APRS gain reflects the value of perturbations; the control gain is bounded even when the upper bounds of the norms of perturbations and initial conditions are unknown.

\noindent \textbf{Notation}. For $\sigma\in \mathbb{R}^m$,  $\Vert \sigma \Vert$ denotes the Euclidean norm. The set $\mathbb{R}_+$ denotes the set of non-negative real numbers. For any square matrix $A$, $\lambda_{\mathrm{min}}(A)$ denotes the smallest eigenvalue of $A$.  Euler method is employed in numerical simulations with sampling step $\Delta_\tau=1\times 10^{-6}$.
\section{Problem statement and main result}
\vspace{-1mm}
Consider a multivariable first order uncertain system
\begin{equation}\label{eq:01}
\begin{aligned}
\dot{\sigma}(t)\!=\!G(t,\!\sigma(t))\![\mathbf{I}\!+\!\Delta g(t,\!\sigma(t))]u(t)\!+\! f(t,\!\sigma(t)),\, \sigma(0)\!=\!\sigma_0,
\end{aligned}
\end{equation} where $\sigma\in \mathbb{R}^{m}$ is the output, $u \in \mathbb{R}^m$ is the control input, $G\in \mathbb{R}^{m \times m}$ is a known function, $f\in \mathbb{R}^{m}$, $\Delta_g \in \mathbb{R}^{m \times m}$ are unknown measurable functions in $t$, for all $\sigma\in \mathbb{R}^m$, and continuous functions in $\sigma$, for almost all $t\geq0$.
\begin{assum}\label{ass0}  For all $(t,\sigma)\in \mathbb{R}_{+}\times \mathbb{R}^m$, $\mathrm{rank}\:G=m$.
\end{assum}
\begin{assum}\label{ass1}  For all $(t,\sigma)\in \mathbb{R}_{+}\times \mathbb{R}^m$, there exist unknown positive constants $d,q>0$ such that $\Vert f(t,\sigma) \Vert \leq d $ and $\Vert G(t,\sigma)\Delta g(t,\sigma)G(t,\sigma)^{-1} \Vert_{\infty}$ $\leq q <1$.
\end{assum}
Consider a control input of the form
\begin{equation}\label{eq:02}
u(t)=G(t,\sigma)^{-1}\nu(t),\: \nu(t)=-\Lambda(t,\sigma)\tfrac{\sigma}{\Vert \sigma \Vert},
\end{equation} where $\Lambda(t,\sigma):\mathbb{R}_+\times \mathbb{R}^m\rightarrow\mathbb{R}_+$ is the controller's gain.  Since \eqref{eq:02} is discontinuous, the solutions of the closed-loop system \eqref{eq:01}-\eqref{eq:02} are understood in the sense of Filippov \cite{filippov13}. Under Assumption \ref{ass1}, any solution of the system \eqref{eq:01}-\eqref{eq:02} with all control components with the same upper-bound $\vert \nu_i(t) \vert \leq C$, $C>0$ satisfies the differential inclusion $\dot{\sigma}_i(t)\in [-d\:,\:d]+[-q\:,\: q]C+\nu_i(t),\: \sigma=0,\:i=1,\ldots,m.$ From this relation it is clear that a sliding mode will be enforced for any sufficiently large gain $C$ since $q<1$.
\begin{assum}\label{ass2}
For all $(t,x)\in \mathbb{R}_{+}\times \mathbb{R}^m$, 
\begin{equation*}
\begin{aligned}
q_1:=\lambda_{\min}\left(\tfrac{1}{2}\right.&(G(t,\sigma)\Delta g(t,\sigma)G(t,\sigma)^{-1} \\
&\left.+G(t,\sigma)^{-T}\Delta g^T(t,\sigma)G(t,\sigma)^T)\right)>-1.
\end{aligned}
\end{equation*}
\end{assum}
\begin{rem}
Assumption \ref{ass2} assures that the uncertain control matrix $\Delta g(t,\sigma)$ could reduce the control effort whenever the eigenvalue $q_1$ takes positive values.
\end{rem}
This paper proposes a solution for the predefined-time RP problem for ASMC still missing to be solved.  In particular, the design of an adaptive gain during RP is presented, ensuring that the RT to a real sliding mode is upper-bounded with a predefined upper-bound.  \vspace{-5pt}
\subsection{Main result}
\vspace{-3mm}
Since upper-bound of perturbations are unknown,  consider the controller's gain of form\vspace{-3pt}
\begin{equation}\label{eq:rpgain}
\begin{aligned}
\Lambda(t,\sigma)&=\hat{\beta}(t)+\kappa(t)\Vert \sigma \Vert,\:  \kappa(t)=\tfrac{1}{\alpha(T_c-t)}, \\
\dot{\hat{\beta}}(t)&=\Vert \sigma \Vert,\: \hat{\beta}(0)=\hat{\beta}_0, 
\end{aligned}
\end{equation} with known positive constants $\alpha\in (0,1)$,  and $T_c>0$ as a prescribed RT upper-bound.  During RP, the gain increases until the value allowing the compensation of the perturbations, such that the system's trajectories reach the set $\left\lbrace \Vert \sigma(t)\Vert = \frac{\varepsilon}{2} \right\rbrace$ in predefined time, i.e., at a time $t=\bar{t}<T_c$ where $\kappa(t)<\infty$ if $\Vert \sigma \Vert \geq \varepsilon /2$ and $\kappa(t)\rightarrow\infty$ as $\sigma \rightarrow 0$. The next lemma resume the main result of the paper, \textit{i.e.},  the predefined RT upper-bound during RP is ensured. Its proof is given in Appendix \ref{app:proofmainlemma}.
\begin{lem}\label{lem:main}
Given any $T_c>0$ and $\varepsilon>0$, $\Vert \sigma_0 \Vert>\varepsilon/2$.  Consider the closed loop system \eqref{eq:01}-\eqref{eq:02} with adaptive gain \eqref{eq:rpgain}. If Assumptions \ref{ass0}-\ref{ass2} are fulfilled, then $\Vert \sigma(\bar{t}) \Vert =\varepsilon/2$ at $t=\bar{t}< T_c$.
\end{lem}
\begin{rem}
The proposed gain in \eqref{eq:rpgain} is composed of two parts.  The proportional term (\textit{cf. }with \cite{gomez2020,holloway2019}) ensures that the norm of the output reaches zero at the prescribed time $T_c$ with unbounded gain. The second part increases more the gain from the beginning,  allowing the output to reach the value $\Vert \sigma(\bar{t}) \Vert = \varepsilon/2$ in a time moment $\bar{t}<T_c$, than only using the first part.  As a result the control objective is ensured with a bounded gain and input.
\end{rem}\vspace{-4pt}
\subsection{Motivating example revisited}\label{ssec:MexampREV}
\begin{figure}[h!]
		\centering
		\includegraphics[width=0.45\textwidth]{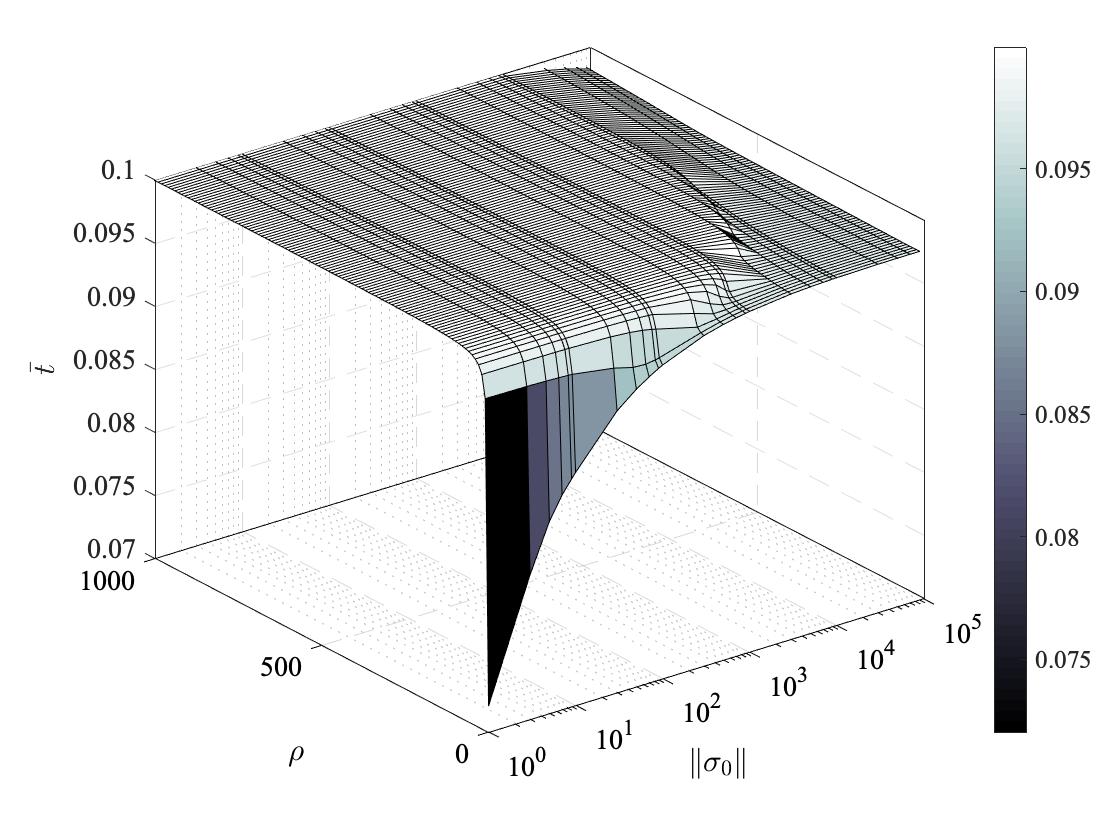}%{figs/prueba3D.jpg}%{figs/BFT1_3D_bone_1200.jpg}
		\caption{Uniform upper-bound of RT under increasing norm of initial conditions and size of perturbations}
		\label{fig:3Duniform}
	\end{figure}

\vspace{-3mm}
Under the same simulation scenario as in the example in subsection \ref{subsec:motexmp},  consider system \eqref{eq:01}-\eqref{eq:rpgain} with an arbitrary RT upper-bound $T_c=0.1$,  $\alpha=0.4$, $\hat{\beta}_0=0$, $
\sigma_0=\frac{b}{\sqrt{2}}\left(
10^{n} \:
-10^{n}
\right)^T,\:G(t,\sigma)=\begin{pmatrix}
2 & -3\\
0 & 3
\end{pmatrix},$
\begin{equation*}
\Delta g(t,\sigma)=\begin{pmatrix}
\tfrac{1}{2}\cos(\sigma_1) & 0.2\cos(\sigma_1)+0.1\sin(5t+\sigma_2)\\
0 & 0.2\cos(\sigma_1)+0.1\sin(5t+\sigma_2)
\end{pmatrix}
\end{equation*} where $\rho \in [0,1000]$, $n\in [1,4]$, $b\in [1,9]$. Figure \ref{fig:3Duniform} illustrates that under the presence of perturbation different size of perturbation and initial conditions,  the trajectories of closed loop system in Lemma \ref{lem:main} will converge to $\left\lbrace\Vert \sigma \Vert = \varepsilon/2 \right\rbrace$ at time $t=\bar{t}< T_c=0.1$. 
This situation enable us to cope the uniform RP strategy to an adaptive design guaranteeing desired properties (i) and (ii).
\section{Case of study: BF based adaptation of SMC }
\vspace{-3mm}
A remarkable approach in ASMC consist in the use of BFs to ensure that a real sliding mode will never be lost without big-overestimation of the perturbation in the ASP.  In this approach, and in order to switch the adaptive gain to a BF, it is required the knowledge of the RT when RP ended. Adopting the RP strategy in Lemma \ref{lem:main},  it is now possible to know when to switch to a BF once the system's trajectories converge into the interior of the $\varepsilon$-vicinity of the sliding manifold at a RT smaller than \textit{a priori} given predefined time. First, we generalize the class of BFs to the multivariable case and then present the complete BF based ASMC approach.
\subsection{Multivariable barrier functions}
\vspace{-3mm}
\begin{defn}\label{def:1} Given $\varepsilon>0$, $\sigma\in \mathbb{R}^{m}$ such that $\Vert \sigma \Vert < \varepsilon$, the multi-variable barrier functions $K_{\mathrm{BF}}(\Vert \sigma \Vert):[0,\varepsilon) \rightarrow [\bar{\beta},\infty)$ are defined as the class of strictly increasing functions in $[0,\varepsilon)$, with vertical asymptote $\lim_{\Vert \sigma  \Vert \rightarrow \varepsilon^{-}} K_{\mathrm{BF}}(\Vert \sigma \Vert)= + \infty$, and a unique global minimum at zero, \textit{i.e.}, $K_{\mathrm{BF}}(0)= \bar{\beta}\geq 0$.
\end{defn}
The class of barrier functions in this paper are those that satisfy the following property. For any positive constant $\beta^{*}\in \mathbb{R}_{+}$, $s:=s(\varepsilon,\bar{\beta},\beta^{*})$ is a root of $K_{\mathrm{BF}}(s)-\beta^{*}=0$ such that $s<\varepsilon$.  Within this class,  in the spirit of \cite{obeid18}, we consider the two types of multi-variable BFs: 
\begin{itemize}
\item Positive definite BF $K_{\mathrm{BF}}(\Vert \sigma \Vert)=K_{pd}(\Vert \sigma \Vert)$ with 
\begin{equation}\label{eq:barrier1}
 K_{pd}(\Vert \sigma \Vert)=\tfrac{\bar{\beta}\varepsilon}{\varepsilon-\Vert \sigma \Vert}, \quad s= \left \{ \begin{matrix} \varepsilon\left(1-\tfrac{\bar{\beta}}{\beta^{*}}\right) & \mathrm{if} & \bar{\beta}<\beta^{*} \\
0 & \mathrm{if} & \bar{\beta}\geq \beta^{*}
\end{matrix}\right.
\end{equation}
\item Positive semi-definite BF $K_{\mathrm{BF}}(\sigma)=k_{psd}(\sigma)$ with
\begin{equation}\label{eq:barrier2}
 K_{psd}(\Vert \sigma \Vert)=\tfrac{\Vert \sigma \Vert}{\varepsilon-\Vert \sigma \Vert}, \quad \bar{\beta}=0,\quad s= \tfrac{\varepsilon \beta^{*}}{1+\beta^{*}}
\end{equation}
\end{itemize}
\begin{rem}
The barrier functions \eqref{eq:barrier1} and \eqref{eq:barrier2} coincide with the ones in \cite{obeid18} for $m=1$.
\end{rem}
\subsection{Refinement of BF based ASMC}
\vspace{-3mm}
Consider the system \eqref{eq:01}-\eqref{eq:02} with adaptive gain
\begin{equation}\label{eq:04} 
\begin{aligned}
\Lambda(t,\sigma)=  \left \{ \begin{matrix} \hat{\beta}(t)+ \kappa(t)\Vert \sigma \Vert,\: \dot{\hat{\beta}}(t)=\Vert \sigma \Vert & \mathrm{if} & 0\leq t < \bar{t}, \\
K_{\mathrm{BF}}(\Vert \sigma \Vert) & \mathrm{if} & t \geq \bar{t},
\end{matrix}\right.
%,\: \hat{\beta}(0)=\hat{\beta}_0, 
\end{aligned} 
\end{equation} where $\bar{t}< T_c$, $\kappa(t):=1/( \alpha(T_c-t))$, with known positive constants $\alpha$,  and $T_c$.  In the first stage, termed RP, the gain increases such that the system's trajectories converge into the manifold $\left\lbrace \Vert \sigma(t)\Vert = \frac{\varepsilon}{2} \right\rbrace$ at $t=\bar{t}<T_c$ despite the size of the upper-bound of perturbation and the initial condition.  During the second stage, termed ASP, the gain is switched to a barrier function that adapts to follow the perturbations variations while ensuring that the trajectories will be contained in an $\varepsilon-$NSS for all future times $t\geq \bar{t}$.  The following result holds whose proof is given in Appendix \ref{app:proofBFtheorem}.
\begin{thm}\label{thm:main}
Given $T_c>0$ and $\varepsilon>0$.  Consider the closed loop system \eqref{eq:01}-\eqref{eq:02} with adaptive gain \eqref{eq:04} If Assumptions \ref{ass1}-\ref{ass2} are fulfilled, then $\Vert \sigma(t) \Vert < s<\varepsilon$ for all $t\geq \bar
{t}$, $\bar{t}\leq T_c$ and any $\sigma_0\in \mathbb{R}^m$.
\end{thm}
\begin{rem}
Notice that the adaptive gain $k(t,x)$ in \eqref{eq:04} switches only once at a time smaller than $T_c$,  without letting that function $\kappa(t)$ grows unbounded. 
\end{rem}
\subsection{Numerical simulation}
\begin{figure}
\centering
\includegraphics[width=0.45\textwidth]{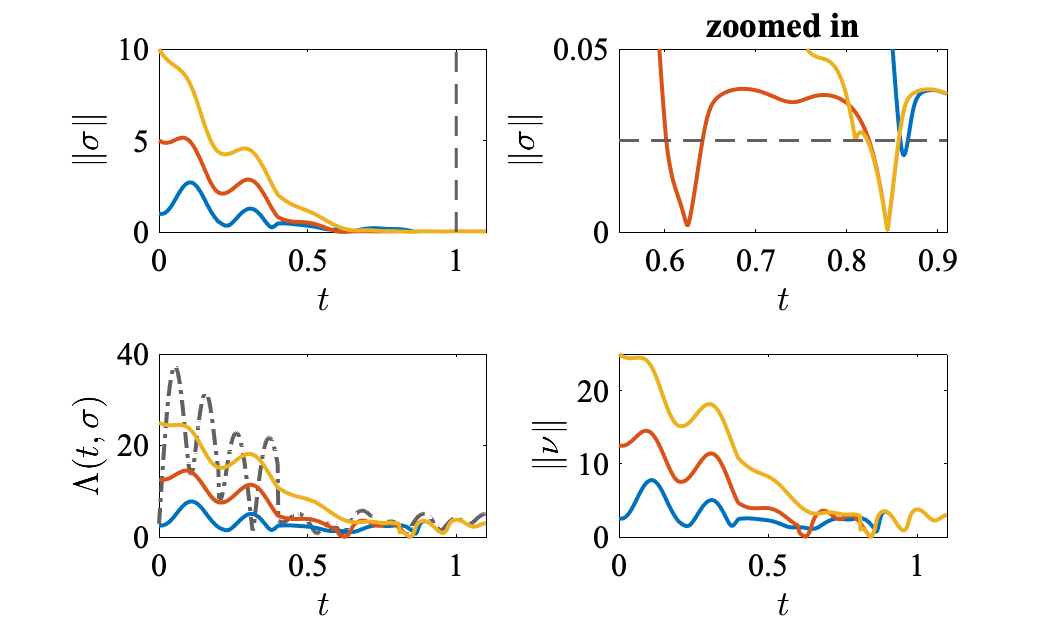}%EX3sc1/ASMC_gainv3.png}
\caption{ARPS+BF scenario 1. (Top-left) Output norm during RP.  (Top-right) Zoomed in output norm at value $\varepsilon/2$. (Bottom-left) adaptive gain (solid curves) vs. Decreasing norm of disturbance (dashed curve). (Bottom-right) Input's norm. }
\label{fig:ex3sc2_decreasing}
\end{figure}

\vspace{-3mm}
{Consider again the motivating example in Section \ref{ssec:MexampREV} and the positive-semidefinte BF in \eqref{eq:barrier2}.  Fix $T_c=0.1$, $\alpha=0.4$, $\hat{\beta}_0=0$, $\varepsilon=0.05$. Two simulation scenarios are illustrated in the presence of bounded disturbances $f_{\rho_1}:=f(t,\sigma,\rho_1)$  and $f_{\rho_2}:=f(t,\sigma,\rho_2)$ with  
\begin{equation*}
\rho_1\!=\!\begin{cases}
          80 &\text{if} \:\: 0\leq t <0.2\\
          50 &\text{if} \:\: 0.2\leq t <0.4\\
          10 & \text{if} \:\: t\geq 0.4
     \end{cases},\rho_2\!=\!\begin{cases}
          10 &\text{if} \:\: 0\leq t <3\\
          100 &\text{if} \:\: 3\leq t <6\\
          200 & \text{if} \:\: t\geq 6
     \end{cases}.
\end{equation*} The first scenario is a closer look at ARPS of Theorem \ref{thm:main} by illustrating in Fig. \ref{fig:ex3sc2_decreasing} the norm of the output, input,  and control gain when a disturbance $\Vert f_{\rho_1} \Vert$ abruptly decreases its value at times $t=0.2$ and $t=0.4$.  Parameters were taken as $\Vert \sigma_0 \Vert \in \left\lbrace 1,5,10 \right\rbrace$, $a=1/\rho_1$, $b=1.2/\rho_1$, $\omega_1=30$, $\omega_2=20$.  As seen in top inset in Fig. \ref{fig:ex3sc2_decreasing}, the output norm attains the value of $\varepsilon/2$ (horizontal dashed line in top-right inset) before time reaches the value of $T_c=1$ (vertical asymptote in top-left inset)  with bounded control gain and input (see bottom insets in Fig. \ref{fig:ex3sc2_decreasing}). For each initial condition,  the solution is continued by switching the control input to the BF at different time instants $\bar{t}$ where $T_c>\bar{t}\in \left\lbrace 0.60182,0.80938,0.85879 \right\rbrace$. Before swiching occurs, the solution does not follow perturbation variations. After switching occurs, the gain becomes lower and then follow perturbation variations with a value less than the norm of the perturbation.}
 %\vspace{-10pt}

\begin{figure}
		\centering
		\includegraphics[width=0.45\textwidth]{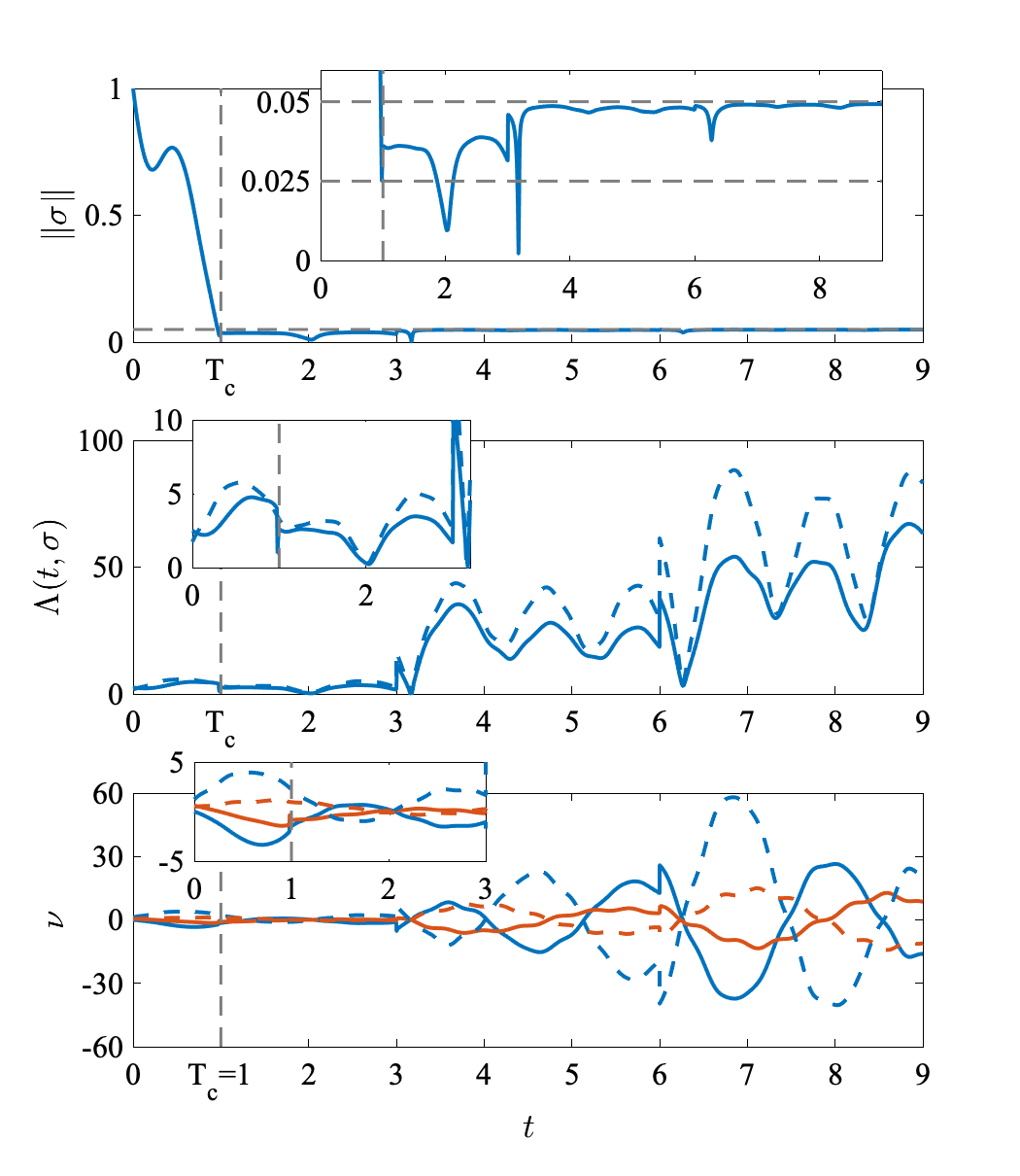}
		\caption{BF+ARPS scenario 2. Trajectories initialized outside of BW (top),  adaptive gain and norm of disturbance (middle),  control signals with corresponding disturbances (bottom). Horizontal dashed asymptotes denote constant values of $\varepsilon=0.05$ and $\varepsilon/2=0.025$, vertical dashed asymptote denotes the prescribed time constant $T_c = 1$, dashed curves denote disturbance norm or signals}
		\label{fig:exp3_comp_outbar}
\end{figure}

{The second scenario consists on the complete illustration of the BF+ARPS approach for a bounded disturbance $f(t,\sigma,\rho)$ with $a=1/\rho_2$, $b=1/\rho_2$, $\omega_1=2$, $\omega_2=3$. Consider $\Vert \sigma_0 \Vert=1$ with $n=0$ and $b=1$ for the symmetric initial conditions in previous examples. Taking initial conditions outside the barrier width (BW) $[0,\varepsilon)$, the norm of the system trajectories during RP in Fig. \ref{fig:exp3_comp_outbar} (top) attains the $\varepsilon/2$ neighbourhood of the sliding set by increasing its gain to reach the disturbance norm before the predefined time convergence (see middle plot in Fig. \ref{fig:exp3_comp_outbar} ). Then,  during ASP, the adaptive gain is switched to a positive semi-definite BF keeping  the norm of trajectories  at lower value than $\varepsilon$ despite that disturbance increases its value at times $t=3,6$.  Notice that the gain is bounded and updating according to disturbance variations while kept at a lower value than the norm of perturbation as illustrated in Fig. \ref{fig:exp3_comp_outbar} (middle). The control signal in Fig. \ref{fig:exp3_comp_outbar} (bottom) is also bounded and continuous (except at the time of switching gain and disturbance), the latter is a consequence of using positive semi-definite BF that decreases towards zero at the same rate than the system trajectories' norm. The behaviour of positive definite BF during ASP can be seen in \cite{obeid18} for $m=1$.}
\section{Conclusion}
\vspace{-3mm}
The ARPS is proposed completing ASMC concept:
\begin{itemize}
\item Controller's gain is adaptive.  The controller gain is adapted in both RP and ASP. Moreover, the upper bound of RT can be predefined in advance. 
\item Adaptive gain is finite.  ASMC+ARPS just require the convergence to the subset $\left\lbrace \Vert \sigma \Vert=\varepsilon /2 \right\rbrace$ that is why it converges in predefined time whose upper-bound is a prescribed time moment.
\item {To show the efficiency of the proposed APRS the BF method is generalized covering two important classes of systems: MIMO systems and systems with uncertain control matrix with unknown upper-bound. It is shown that a combination of proposed ARPS with BF adaptation ensure a predefined time convergence to NSS; the APRS gain reflects the value of perturbations; the control gain is bounded even when the upper bounds of the norms of perturbations and initial conditions are unknown. APRS can be extended after the moment when a solution will reach the set $\left \lbrace \Vert \sigma \Vert= \varepsilon/2\right \rbrace$ if the gain is switched to any other ASMC algorithm, not restricted directly to other discontinuous sliding mode algorithms or continuous ones with appropriate modifications.}
\end{itemize}
\vspace{-2mm}
\begin{ack}      
\vspace{-3mm}                     
This research was supported by DGAPA-UNAM (Programa de Becas Posdoctorales DGAPA en la UNAM), CONACyT (Consejo Nacional de Ciencia y Tecnolog\'ia), Project 282013; PAPIIT--UNAM (Programa de Apoyo a Proyectos de Investigaci\'on e Innovaci\'on Tecnol\'ogica) IN115419.
\end{ack}

\bibliographystyle{plain}        % Include this if you use bibtex 
\bibliography{ASMC_rp_arxiv}           % and a bib file to produce the 
\appendix 
\vspace{-3mm}
\section{Proof of Lemma \ref{lem:main}}\label{app:proofmainlemma}\vspace{-10pt}
By using a time scale transformation,  it is shown that RP ends before a prescribed convergence time $T_c$ (uniformly in the initial conditions and upper-bound of perturbations).  Consider the uncertain system \eqref{eq:01} and the time scale transformation given in \cite{gomez2020}\vspace{-5pt}
\begin{equation}\label{eq:time_scalling}
 t=T_c\left(1-\e^{-\alpha \tau}\right) \: \Leftrightarrow \:\tau=-\alpha^{-1} \ln\left(1-\tfrac{t}{T_c}\right)
\end{equation} with $\alpha, \: T_c>0$, the resulting time scaled system is
\begin{equation}\label{eq:tssys}
\begin{aligned}
y^{'}(\tau)&=[(\mathbf{I}+\overline{\Delta G}(\tau,y))\bar{\kappa}(\tau)^{-1}\nu(\tau) +\bar{ f}(\tau,y(\tau))], 
\end{aligned}
\end{equation} where  $y(\tau):=\sigma(t)\vert_{\eqref{eq:time_scalling}}$ is the state,  $\nu(\tau):=u(t)\vert_{\eqref{eq:time_scalling}}$ is the input,  $\overline{\Delta G}(\tau,y):={G}(t,\sigma){\Delta g}(t,\sigma){G}(t,\sigma)^{-1}\vert_{\eqref{eq:time_scalling}}$ satisfies Assumption \ref{ass2} and $y^{'}:=\dfrac{\d y}{\d \tau}$. The time scaling makes the perturbation to vanish with $\bar{\kappa}(\tau)^{-1}:=\kappa(t)^{-1}\vert_{\eqref{eq:time_scalling}}=\alpha T_c \e^{-\alpha \tau}$, i.e,   
\begin{equation}\label{eq:vanishingpert}
\bar{f}(\tau,y(\tau)):=\bar{\kappa}(\tau)^{-1} f(t,x(t))\vert_{\eqref{eq:time_scalling}}
\end{equation} as time $\tau$ grows unbounded. From \eqref{eq:01},  \eqref{eq:02} and \eqref{eq:rpgain} the control law is given as follows
\begin{equation}\label{eq:tscontrol}
\begin{aligned} 
 \nu(\tau)=-\hat{\tilde{\beta}}(\tau)\tfrac{y}{\Vert y \Vert}-{y},\: \hat{\tilde{\beta}}^{'}(\tau)&=\bar{\kappa}(\tau)^{-1}\Vert {y} \Vert
\end{aligned}
\end{equation} where $\hat{\tilde{\beta}}(\tau):=\hat{\beta}(t)\vert_{\eqref{eq:time_scalling}}$.  
Next we prove that $y(\tau)$ converges to the manifold $\Vert y(\tau) \Vert=0$ as $\tau$ grows unbounded, this means that $\sigma(t)$ converges to $\Vert \sigma(t) \Vert=0$ as $t\rightarrow T_c$.
Consider the Lyapunov function $V(\tau)=V_1(\tau)+V_2(\tau)$,
\begin{equation}\label{eq:lyap_fun}
\begin{aligned}
V_1(\tau)&=\Vert y \Vert^2,\:V_2(\tau)=b_0(\hat{\tilde{\beta}}(\tau)-\beta^*)^2,
\end{aligned}
\end{equation} where $\beta^*:=d/b_0$ and $b_0:=(1+q_1)$ are unknown positive constants, $q_1>-1$ as in Assumption \ref{ass2}. The time derivative of $V(\tau)$ gives $V^{'}(\tau)=V_1^{'}(\tau)+V_2^{'}(\tau)$:
\begin{itemize}
\item For $V_1^{'}(\tau)$, by using  \eqref{eq:tssys} and \eqref{eq:tscontrol} it holds that
\begin{equation*}
\begin{aligned}
V_1^{'}(\tau)&=2{y}^T(-\hat{\tilde{\beta}}(\tau)\bar{\kappa}^{-1}(\tau)\tfrac{y}{\Vert y \Vert}-y)+2{y}^T\bar{f}(\tau,y)\\
&+2{y}^T\overline{\Delta G}(\tau,y)(-\hat{\tilde{\beta}}(\tau)\bar{\kappa}(\tau)^{-1}\tfrac{y}{\Vert y \Vert}- {y})
\end{aligned}
\end{equation*} By using the fact that ${y}^T\overline{\Delta G}{y}={y}^T\overline{\Delta G}^T{y}$ and Cauchy-Schwartz inequality, it holds that
\begin{equation*}
\begin{aligned}
&V_1^{'}\leq-2\hat{\tilde{\beta}}(\tau)\bar{\kappa}(\tau)^{-1}\Vert {y}\Vert-2 \Vert {y}\Vert^2+2\Vert {y}\Vert \Vert \bar{f}(\tau,y) \Vert\\
&-2\hat{\tilde{\beta}}(\tau)\bar{\kappa}(\tau)^{-1}y^T(\tfrac{1}{2}(\overline{\Delta G}(t,y)+\overline{\Delta G}^T(t,y)))\tfrac{y}{\Vert y \Vert}\\
&-2y^T(\tfrac{1}{2}(\overline{\Delta G}(t,y)+\overline{\Delta G}^T(t,y))){y}
\end{aligned}
\end{equation*} It follows from \eqref{eq:vanishingpert} and Assumption \ref{ass1} that $\Vert \bar{f} \Vert \leq \bar{\kappa}(\tau)^{-1}d$, hence
\begin{equation}\label{eq:V1bound}
\begin{aligned}
V_1^{'}(\tau)&\leq -2\hat{\tilde{\beta}}(\tau)\bar{\kappa}(\tau)^{-1}\Vert {y} \Vert-2\Vert {y}\Vert^2+2d\bar{\kappa}(\tau)^{-1} \Vert {y} \Vert\\
&-2q_1\hat{\tilde{\beta}}(\tau)\bar{\kappa}(\tau)^{-1}\Vert {y} \Vert-2q_1\Vert y \Vert^2\\
&= -2b_0\bar{\kappa}(\tau)^{-1}(\hat{\tilde{\beta}}(\tau)-\beta^{*})\Vert {y} \Vert-2 b_0\Vert {y} \Vert ^2
\end{aligned}
\end{equation} where we used the Rayleigh-Ritz inequality.
\item For $V_2^{'}(\tau)$,  taking into account \eqref{eq:tscontrol}-\eqref{eq:lyap_fun},  it holds
\begin{equation}\label{eq:V2bound}
\begin{aligned}
V_2^{'}(\tau)\!=\!2b_0(\hat{\tilde{\beta}}(\tau)\!-\!\beta^{*})\hat{\tilde{\beta}}^{'}\!\!(\tau)\!=\!2b_0\bar{\kappa}(\tau)^{\!-\!1}\!(\hat{\tilde{\beta}}(\tau)\!-\!\beta^{*})\Vert {y} \Vert
\end{aligned}
\end{equation}
\end{itemize} \vspace{-10pt}By setting $W(y):=-2b_0\Vert y \Vert^2$ and using \eqref{eq:V1bound} and \eqref{eq:V2bound} it holds that $V^{'}(\tau)\leq -W(y)\leq 0$. This implies that $V(\tau)\leq V(0)$ and $y$, $\hat{\tilde{\beta}}$ are bounded.  On the one hand $\int_0^\infty W(y)\mathrm{d}s\leq V_0-V(t)<\infty$. Moreover, $W(y)$ is continuous and since $y(\tau)$ is bounded and uniformly continuous (its derivative is bounded from Assumption \ref{ass1} and \eqref{eq:tssys}), then $W(y)$ is also uniformly continuous. From Barbalat's Lemma \cite{khalil2001}, then $W(y)\rightarrow 0$ as $\tau \rightarrow \infty$, which implies that $y(\tau)$ converges to the set $\left\lbrace \Vert y \Vert=0 \right\rbrace$ as $\tau$ grows unbounded. Since $y$ converges asymptotically to zero, there exist a function $\eta \in \mathcal{KL}$ such that for $\tau\geq 0$
\begin{equation}\label{eq:Tepsp2}
\Vert y \Vert \leq \eta(\Vert y_0 \Vert,\tau)\leq \eta(c,\tau),
\end{equation} due to $y$ is bounded with $c>0$. Finally, since $\eta(c, \tau)\rightarrow 0$ as $\tau$ grows unbounded, given $\varepsilon>0$ there exists $\tau_0>0$ such that $\eta(c,\tau)< \varepsilon/2$,  whenever $\tau \geq \tau_0$. Then, take $\bar{\tau}\geq \tau_0$ and from \eqref{eq:Tepsp2} it follows that $\Vert y(\tau) \Vert < \varepsilon/2$ for all $\tau\geq \bar{\tau}$.  Equivalently, by means of the time-scaling \eqref{eq:time_scalling},  $\Vert \sigma(t) \Vert =\varepsilon/2$ in a time $\bar{t}=\lim_{\tau\rightarrow\bar{\tau}}t:=\lim_{\tau\rightarrow\bar{\tau}}T_c(1-\e^{-\alpha \tau})<T_c$, where $T_c$ is an arbitrary \textit{a priori} given constant independent of the initial condition and the upper-bound of perturbations.  \vspace{-10pt}
\section{Proof of Theorem \ref{thm:main}}\label{app:proofBFtheorem} 
\vspace{-3mm}
Following the proof of Lemma \ref{lem:main}, it is ensured that the system's trajectories reach the value $\Vert \sigma(t) \Vert \leq \varepsilon/2$ at time $t=\bar{t}<T_c$. Then, it is left to prove that system’s trajectories will be contained in a region $\Vert \sigma(t) \Vert < \varepsilon$ for all future times $t\geq \bar{t}$.

Let $t=\bar{t}$ denote the first time such that $\Vert \sigma(t) \Vert \leq \varepsilon/2$ and consider the barrier functions given in \eqref{eq:barrier1}-\eqref{eq:barrier2}. The result follows from using the next auxiliary lemma, which is the generalization to the multivariable case of the barrier function based ASMC.
\vspace{-1mm}
\begin{lem}\label{lem:barrierfun}
Consider that Assumptions \ref{ass0}-\ref{ass2} are fulfilled. Given the uncertain system \eqref{eq:01} controlled by \eqref{eq:02} with $k(t,\sigma)=K_{\mathrm{BF}}(\Vert \sigma \Vert)$. Then, for all $t \geq \bar{t}$ and for all $\Vert \sigma(t) \Vert > s$,  the sliding variable $\sigma(t)$ converges in finite time to a region $\left\lbrace \Vert \sigma(t)\Vert \leq s < \varepsilon \right\rbrace$.
\end{lem}\vspace{-10pt}
\begin{pf}
Consider the closed loop system \eqref{eq:01}-\eqref{eq:02} and adaptive gains as a barrier function, 
\vspace{-1mm}
\begin{equation}\label{eq:A1}
\begin{aligned}
\dot{\sigma}(t)&=-(\mathbf{I}+\Delta G(t,\sigma))K_{\mathrm{BF}}(\Vert \sigma \Vert)\tfrac{\sigma}{\Vert \sigma \Vert}+f(t,\sigma)\\
\end{aligned}
\end{equation} where $\Delta G(t,\sigma):=G(t,\sigma)\Delta g(t,\sigma)G(t,\sigma)^{-1}$. Notice %that %the time derivative of the class of barrier functions \eqref{eq:barrier1} or \eqref{eq:barrier2} yields to 
\begin{equation}\label{eq:A2}
\dot{K}_{\mathrm{BF}}(\Vert \sigma \Vert)=\tfrac{\theta \varepsilon}{(\varepsilon-\Vert \sigma \Vert)^2}\tfrac{\sigma^T\dot{\sigma}}{\Vert \sigma \Vert}
\end{equation} with the convention that $\theta=1$ if $K_{\mathrm{BF}}=K_{\mathrm{psd}}$ or $\theta=\bar{\beta}$ if $K_{\mathrm{BF}}=K_{\mathrm{pd}}$.  Consider the Lyapunov function $V(t)=\tfrac{1}{2}\Vert \sigma \Vert^2+\tfrac{1}{2}(K_{\mathrm{BF}}(\Vert \sigma \Vert)-K_{\mathrm{BF}}(0))^2$. The time derivative of $V(t)$ along the trajectories of \eqref{eq:A1}-\eqref{eq:A2} is given by
\vspace{-1mm}
\begin{equation*}
\begin{split}
\dot{V}(t)&\!=\!-K_{\!\mathrm{BF}\!}(\Vert \sigma \Vert)\Vert \sigma \Vert-K_{\!\mathrm{BF}\!}(\Vert \sigma \Vert) \sigma^T\Delta G(t,\sigma)\tfrac{\sigma}{\Vert \sigma \Vert}\\
&+\sigma^Tf(t,\sigma)-K_{\!\mathrm{BF}\!}(\Vert \sigma \Vert)\zeta(K_{\!\mathrm{BF}\!}(\Vert \sigma \Vert)
\!-\!K_{\!\mathrm{BF}\!}(0))\\
&\!-\!K_{\!\mathrm{BF}\!}(\Vert \sigma \Vert)\zeta(K_{\!\mathrm{BF}\!}(\Vert \sigma \Vert)\!-\!K_{\!\mathrm{BF}\!}(0))\!\tfrac{\sigma^T}{\Vert \sigma \Vert}\Delta G(t,\sigma)\tfrac{\sigma}{\Vert \sigma \Vert}\\
&+\zeta(K_{\!\mathrm{BF}\!}(\Vert \sigma \Vert)\!-\!K_{\!\mathrm{BF}\!}(0))\tfrac{\sigma^T}{\Vert \sigma \Vert}f(t,\sigma),
\end{split}
\end{equation*} where $\zeta:=\theta \varepsilon/(\varepsilon-\Vert \sigma \Vert)^2$, $\theta \in \left\lbrace  1, \bar{\beta} \right \rbrace$. By using the Cauchy-Schwarz inequality, Rayleigh-Ritz inequality and Assumption \ref{ass1}, the following upper bound holds
\vspace{-2mm}
\begin{equation}\label{eq:A4}
\begin{aligned}
\dot{V}(t)&\leq -b_0(K_{\!\mathrm{BF}\!}(\Vert \sigma \Vert)\!-\!\beta^{*})\Vert \sigma \Vert\\
&\!-\!b_0\zeta(K_{\!\mathrm{BF}\!}(\Vert \sigma \Vert)\!-\!\beta^{*})\vert K_{\!\mathrm{BF}\!}(\Vert \sigma \Vert)\!-\!K_{\!\mathrm{BF}\!}(0)\vert\\
&=-b_0\beta_s\Vert \sigma \Vert-b_0\zeta \beta_s\vert K_{\mathrm{BF}}(\Vert \sigma \Vert)-K_{\mathrm{BF}}(0)\vert,
\end{aligned}
\end{equation} where $\beta_s:=K_{\mathrm{BF}}(\Vert \sigma \Vert)-\beta^{*}$, $\beta^{*}=d/b_0$, $b_0:=1+q_1$ and we used the fact that $\sigma^T \Delta G(t,\sigma) \sigma=\sigma^T \Delta G(t,\sigma)^T \sigma$. Following similar arguments as in \cite{obeid18}, the following three cases are considered:
\vspace{-1mm}
\begin{itemize}
\item[(i)] Let $\bar{\beta}<\beta^{*}$ when $\Vert \sigma \Vert > s$ with $s$ defined as in \eqref{eq:barrier1} or \eqref{eq:barrier2}. Since barrier functions are strictly incresing functions in $\Vert \sigma \Vert$,  then $K_{\mathrm{BF}}(\Vert \sigma \Vert)>K_{\mathrm{BF}}(s)=\beta^{*}$ on $s<\Vert \sigma \Vert<\varepsilon$. Hence $\beta_s>0$ and 
\begin{equation*}
\begin{split}
\dot{V}&\leq -b_0\beta_s\min \left\lbrace1,\: \zeta \right\rbrace(\tfrac{\Vert \sigma \Vert}{\sqrt{2}}+\tfrac{\vert K_{\mathrm{BF}}(\Vert \sigma \Vert)-K_{\mathrm{BF}}(0) \vert}{\sqrt{2}})\\
&\leq -\beta_0V^{\tfrac{1}{2}}, \: \beta_0:=b_0\beta_s\min \left\lbrace1,\: \zeta \right\rbrace.
\end{split}
\end{equation*}
\item[(ii)] Let $\bar{\beta}\geq \beta^{*}$ when $\Vert \sigma \Vert > s$ as in \eqref{eq:barrier1}. Then, $K_{\mathrm{BF}}\!\!>\!K_{\mathrm{BF}}(0)\!\!=\!\!\bar{\beta}\!\!\geq \!\! \beta^{*}$. From \eqref{eq:A4}, it holds that $\dot{V}\leq -\beta_0 V^{1/2}$.
\item[(iii)] For $\Vert \sigma \Vert < s$, \eqref{eq:A4} would be sign indefinite until the solution $\sigma(t)$ reaches the set $\left\lbrace \Vert \sigma \Vert=s \right\rbrace$ and $\dot{V}\leq 0$ as $\beta_s=0$.  Hence, $V$ remains constant or decreasing. This implies that $\Vert \sigma \Vert \leq s$ for all times.
\end{itemize} Let $\Omega_1\!=\!\left \lbrace \Vert \sigma\Vert\leq s \right \rbrace$ and $\Omega_2\!=\!\left \lbrace s<\Vert \sigma\Vert< \varepsilon \right \rbrace$. Items (i) and (ii) ensure finite time convergence to the domain $\Omega$ for all $t\geq \bar{t}+T$ if the solution starts in $\Omega_2$. Finally, by construction $s<\varepsilon$ (see \eqref{eq:barrier1} and \eqref{eq:barrier2}), and it follows from (iii) that $\Vert \sigma \Vert \leq s < \varepsilon$ holds for all time $t\geq \bar{t}+T$. If the solution starts in $\Omega_1$, the same result follows from (iii) with $T=0$. \hfill $\square$
\end{pf}
\end{document}